\documentclass[aps,prb,superscriptaddress,showpacs,amsmath,twocolumn]{revtex4}

\usepackage{amsmath}
\usepackage{amssymb}
\usepackage{graphicx}

\begin{document}


\title{Notes on the minimal longitudinal dc conductivity of perfect 
bilayer graphene}

\author{ J{\'o}zsef Cserti
}
\affiliation{Department of Physics of Complex Systems,
E{\"o}tv{\"o}s University
\\ H-1117 Budapest, P\'azm\'any P{\'e}ter s{\'e}t\'any 1/A, Hungary}



\begin{abstract}

We calculated the minimal longitudinal conductivity in prefect single
and bilayer graphene by extending the two methods developed 
for Dirac fermion gas by A. W. W. Ludwig et al.\ in 
Phys. Rev. B {\bf 50}, 7526 (1994). 
Using the Kubo formula which was originally applied for spintronic
systems we obtain 
$\sigma^{\text min}_{xx}= (J \pi /2)\, e^2/h$
while from the other formula used in the above mentioned work we find 
$\bar{\sigma}^{\text min}_{xx}= (4J/\pi)\, e^2/h$, where 
$J=1$ for single layer and $J=2$ for bilayer graphene. 
The two universal values are different although they are 
numerically close to each other. 
Our two results are in the same order of magnitude as that of
experiments and for single layer case one of our result agrees 
many earlier theoretical predictions. 
However, for bilayer graphene only two studies are known with
predictions for the minimal conductivity different from our
calculated values. 
Similarly to the single layer case, the physical origin of 
the minimal conductivity in bilayer graphene is also 
rooted back to the intrinsic disorder induced by the Zitterbewegung which
is related to the trembling motion
of the electron.

\end{abstract}

\pacs{81.05.Uw, 73.23.Ad, 72.10.Bg, 73.43.Cd}

\maketitle


Unusual and remarkable transport properties of graphene 
(single or stacks of atomic layer of graphite) have been 
proved by recent experiments\cite{Novoselov-1:cikk,Kim:cikk}. 
Besides the unconventional quantum Hall effect, many new phenomena
have been predicted and studied in graphene 
such as the Klein paradoxon\cite{Katsnelson-Klein:cikk}, 
the specular Andreev 
reflection\cite{Carlo-Andreev_graphene:cikk,Bhattacharjee:cikk}, 
the Josephson effect\cite{Carlo-Josephson:cikk}, 
a new electric field effect\cite{Lukose:cikk}, 
the photon-assisted electron transport\cite{Trauzettel:cikk}, 
composite Dirac fermions\cite{Khveshchenko-composit_graphene:cikk},      
quantum dots\cite{Silvestrov_graphene_dot:cikk}, 
the n-p junction\cite{Falko_n-p-junction:cikk}, 
the fractional quantum Hall effect\cite{Toke_fractional_QHE:cikk},  
the spin-orbit gap\cite{Yao_spin-orbit_gap:cikk}. 

In this work we study the longitudinal conductivity of graphenes
which, according to experiments, 
takes the minimum values of orders of $e^2/h$. 
This is an intrinsic property in the sense that it persists 
even in perfect (impurity free) carbon honeycomb lattice.
Such a peculiar behavior can be related to the excitation spectrum
of single layer graphene described well 
by a Dirac like dispersion relation (Dirac cones) 
of two-dimensional massless Dirac fermions\cite{McClure_DiVincenzo:cikk}. 

Much theoretical efforts have been devoted to explain
quantitatively the observed minimal longitudinal conductivity. 
However, at the theoretical side different predictions exist for the
value of the minimal conductivity $\sigma^{\text min}_{xx}$ 
even in prefect single layer graphene. 
Actually, long before the experimental evidence of the minimal
conductivity, it has already been considered 
theoretically\cite{Fradkin:cikk,Lee:cikk,Gusynin-1:cikk} 
and Ludwig et al.\ found different
values using two different approaches\cite{Ludwig:cikk}.  
Similarly, in many recent 
works\cite{Gusynin-Hall_sxx:cikk,Peres-1:cikk,Katsnelson:cikk,Tworzydlo:cikk} 
it has been found that $\sigma^{\text min}_{xx} = (4/\pi)\, e^2/h$, 
while Ziegler predicted\cite{Ziegler-robust:cikk} 
$\sigma^{\text min}_{xx} = \pi \, e^2/h$, 
Falkovsky and Varlamov\cite{Falkovsky:cikk} obtained 
$\sigma^{\text min}_{xx} = (\pi/2)\, e^2/h$, 
and Nomura and MacDonald obtain from numerical calculations of the Kubo
formula $\sigma^{\text min}_{xx} = (4/\pi)\, e^2/h$ for short range 
scattering case and $\sigma^{\text min}_{xx} =  4\, e^2/h$
for Coulomb scattering case\cite{Nomura_MacDonald:cikk}.
Recently,  Gusynin and Sharapov have derived analytical results 
for the ac and dc conductivity of Dirac fermions 
in graphene\cite{Gusynin-Dirac-transport:cikk} 
which could be used for studying the minimal conductivity.

For bilayer graphene first studied experimentally\cite{Novoselov-bilayer:cikk} 
by Novoselov et al.\  and theoretically\cite{Ed-Falko:cikk} 
by McCann and  Fal'ko, the situation is not better regarding 
the minimal conductivity.
Much less theoretical works considered the minimal conductivity 
in bilayer graphene.  
Recently, Koshino and Ando have investigated the transport in bilayer
graphene in self-consistent Born approximation\cite{Ando-bilayer:cikk} 
and they found that in the strong-disorder regime 
$\sigma^{\text min}_{xx} = (8/\pi)\, e^2/h$, while in the weak-disorder
regime $\sigma^{\text min}_{xx} = (24/\pi)\, e^2/h$ which is six times
larger than in single layer graphene.
Similarly, Katsnelson has also calculated the minimal conductivity in
bilayers using the Landauer 
approach\cite{Ando-bilayer:cikk,Katsnelson-bilayer:cikk} 
and he obtained a different value 
$\sigma^{\text min}_{xx} = 2\, e^2/h$.  

For massless Dirac fermion (corresponding to the single layer
graphene) Ludwig et al.\ calculated the conductivity\cite{Ludwig:cikk} 
using two definitions for that, the one is  
the Kubo formula ($\sigma^{\text min}_{xx}$) 
and the other ($\bar{\sigma}^{\text min}_{xx}$) 
is a definition often used in the sigma model
literature\cite{McKane:cikk}. 
The two definitions yield two different results 
(although numerically they are close to each other) 
for the longitudinal conductivity of perfect single layer graphene, 
namely, $\sigma^{\text min}_{xx} = n_v n_s (\pi/8) \, e^2/h$ and 
$\bar{\sigma}^{\text min}_{xx} = n_v n_s (1/\pi) \, e^2/h$, 
respectively. Here the factor $n_v=2$ and $n_s = 2$ correspond 
to the two valleys and the electron spin, respectively (for details
see references before). 

As can be seen from the above reviewed literature, several  
predictions of the exact value of the
minimal conductivity even in perfect single and bilayer graphene exist
although they are consistent at least in order of magnitude.  
In this work we extend Ludwig et al.\ approach to calculate 
the longitudinal conductivity for bilayer graphene. 
We find that, similarly to the case of single layer graphene, 
the two definitions of the conductivity used in Ref.~\onlinecite{Ludwig:cikk} 
yield also different results for perfect bilayer graphene. 
More interestingly, we find that the conductivity obtained from 
both approaches is {\em doubled} compared to that for single layer graphene. 
Thus, our results show that obtaining the exact value of the minimal
conductivity is a rather subtle task and it needs further
investigation. 

Regarding the technical details of our calculations we note that 
when we used the Kubo formula, instead of
following the approach of Ludwig et al. we applied an {\em equivalent} 
method developed by Bernevig~\cite{Bernevig:cikk} for spintronic systems. 
In this method the conductivity is calculated in `bubble'
approximation using the one-particle non-interacting Green's function 
for finite temperature.
This approach results in the same conductivity for single layer graphene
as that obtained from Ludwig et al. method and, as can be seen below, 
it is more convenient to extend to bilayer graphene. 

In order to treat the single and bilayer graphene simultaneously we
start from the Hamiltonian given in a unified form  
as~\cite{Novoselov-1:cikk,Ed-Falko:cikk}
\begin{eqnarray}
H_J &=& g \,  \left( \begin{array}{cc}
0 & {\left(p_x - i p_y\right)}^J \\
{\left(p_x + i p_y\right)}^J & 0
\end{array}  \right), 
\label{HJ-1:eq}
\end{eqnarray}%
where $J=1$ for single and $J=2$ for bilayer graphene, 
and $g$ is a constant depending on $J$. 
In what follows, it is more convenient to use the following equivalent
form of Eq.~(\ref{HJ-1:eq})
\begin{eqnarray}
H_J &=& \mbox{\boldmath $\Omega$}({\bf p}) \,\mbox{\boldmath $\sigma$},  
\label{HJ_omega:eq}
\end{eqnarray}%
where $\mbox{\boldmath $\sigma$} = (\sigma_x,\sigma_y)$ 
are the Pauli matrices representing the `pseudo spin' acting on the
spinor states with components corresponding to the wave function's
amplitudes at the two non-equivalent basis atoms in the unit cell of
the honeycomb lattice,  
$\mbox{\boldmath $\Omega$} ({\bf p}) = 
g\, p^J (\cos J\varphi, \sin J\varphi)$ and 
$\varphi$ is the polar angle of ${\bf p}$, i.e., 
${\bf p}= p\, (\cos \varphi, \sin \varphi)$ 
with $p= \left|{\bf p}\right|$. 
The eigenstates of the Hamiltonian $H_J$ with plane wave solutions 
$e^{i {\bf k}{\bf r} }$ have eigenvalues 
\begin{eqnarray}
E_\pm ({\bf k}) &=& \pm \Omega ({\hbar \bf k}),
\label{Epm:eq}
\end{eqnarray}%
where $\Omega = \sqrt{{{\mbox{\boldmath $\Omega$}^2}({\hbar \bf k})}}$ is
the magnitude of the vector ${\mbox{\boldmath $\Omega$}({\hbar \bf k})}$. 

The Green's function of a Hamiltonian $H$ is defined by 
$\hat{G} (z) = {(z - \hat{H})}^{-1}$ 
and its position representation reads 
$ G({\bf r},{\bf r}^\prime,z) 
= \langle {\bf r} | \hat{G}(z) | {\bf r}^\prime \rangle $. 
Using the Hamiltonian (\ref{HJ_omega:eq}) one finds 
\begin{eqnarray}
\lefteqn{G({\bf r},{\bf r}^\prime,z) 
= \langle {\bf r} | G (z)| {\bf r}^\prime \rangle } \nonumber \\[2ex]
&=& \int \frac{d^2 k}{{\left(2\pi \right)}^2}\, 
e^{i{\bf k }\left({\bf r}-{\bf r}^\prime \right)} \, 
\frac{ z
+ \mbox{\boldmath $\Omega$}({\bf k}) \,\mbox{\boldmath $\sigma$}}
{z^2 - {\mbox{\boldmath $\Omega$}^2({\bf k})}},
\label{Green:eq}
\end{eqnarray}
where $z$ is a complex number. 

To calculate the conductivity from the Kubo formula we apply the result
derived first by Bernevig~\cite{Bernevig:cikk} for spintronic
systems using the Green's function (\ref{Green:eq}): 
\begin{subequations}
\begin{eqnarray}
\sigma_{ij}(\omega) &=& 
-\frac{e^2}{\hbar}\, 
\lim_{\omega \to 0}
\frac{\text{Im}\{\Pi^{\text{ret}}_{ij}(\omega)\}}{\omega}, \,\,\, 
\text{where} 
\label{cond-Kubo:eq} \\[2ex]
\Pi_{ij}(\omega) &=& \frac{2}{A}\, \sum_{{\bf k}}
\frac{n_F(E_+)-n_F(E_-)}{\omega^2 - 4\Omega^2}\, \Omega^2 
\nonumber \\[2ex]
&\times& 
\left(\, i \omega\, \epsilon_{\alpha \beta \gamma} 
F_{i\alpha} F_{j\beta}\, \hat{\Omega}_\gamma 
- 2 \Omega \, F_{i\alpha} F_{j\alpha} \right), 
\label{Pi:def}  \\[2ex]  
F_{i\alpha} &=& 
\frac{\partial \hat{\Omega}_\alpha}{\partial k_i}, 
\,\,\,\, \hat{\Omega}_\alpha = \frac{\Omega_\alpha}{\Omega}, 
\end{eqnarray}%
\label{Kubo_cond:eq}
\end{subequations}%
and $\Pi^{\text{ret}}_{ij}(\omega)$ is the retarded correlation
function obtained by analytic continuation $\omega \to \omega +
i\eta$ of function $\Pi_{ij}(\omega)$,  
$n_F(E)=1/(e^{\beta (E-\mu)}+1)$ is the Fermi function 
with Fermi energy $\mu$, the indices take the values $1,2,3$, 
and summation over repeated indices is assumed.
Here $A$ is the area of the two dimensional system, 
$\epsilon_{\alpha \beta \gamma}$ is the Levi-Civita symbol, and   
the two energy bands $E_{\pm}({\bf k})$ are given by Eq.~({\ref{Epm:eq}). 
This result is valid for ballistic systems (note that this
result is slightly rewritten to make it more transparent and useful
for further calculations).

We now evaluate the conductivity for undoped single and bilayer
graphene at zero temperature using Eq.~(\ref{Kubo_cond:eq}). 
The Fermi energy for undoped graphene
is $\mu = 0$, i.e., the negative energy band is fully occupied and the
positive one is empty at zero temperature.
Thus, in Eq.~(\ref{Pi:def})  $n_F(E_+)-n_F(E_-) =-1$ for all occupied states
with Fermi wave number ${\bf k}$.   
For dc conductivity we need to find $\sigma(\omega)$ in
the limit $\omega \to 0 $. 
Therefore, the integration over ${\bf k}$ can safely be extended 
to infinity since the contribution comes from the integrand 
close to $k\approx 0$ (see the denominator of $\Pi_{ij}(\omega)$). 
Using Eq.~(\ref{Pi:def}) and $\sum_{\bf k} \to A \int d^2 k /(2\pi)^2$ 
(no contribution comes from the term contains 
the Levi-Civita symbol $\epsilon_{\alpha \beta \gamma}$) 
we find  
\begin{eqnarray}
\Pi^{\text{ret}}_{xx}(\omega)  &=& \lim_{\eta \to 0^+}\, 
\int_{0}^\infty
\frac{dk}{2 \pi}\, 
\frac{2g^3 J^2 k^{3 J-1}}{{(\omega + i\eta)}^2-4g^2 k^{2J}}.  
\end{eqnarray}%
The imaginary part of the above integral can be evaluated analytically
and it yields  
\begin{eqnarray}
\text{Im}\{\Pi_{xx}(\omega)\} &=& -\frac{J\omega}{16}.
\end{eqnarray}%
Thus, from (\ref{cond-Kubo:eq}) the longitudinal dc conductivity 
(per valley per spin) for single ($J=1$) and bilayer ($J=2$) graphene 
becomes: 
\begin{eqnarray}
\sigma_{xx} &=& \frac{J\pi}{8}\, \frac{e^2}{h}. 
\label{sigma__per-valley_per-spin:eq}
\end{eqnarray}%
Note that the conductivity obtained from the Kubo formula 
is universal depending only on the type of
graphene but not the constant $g$. 

We now apply the other definition used by Ludwig et al.\ for
calculating the conductivity (see Eq.~(55) 
in Ref.~\onlinecite{Ludwig:cikk}) which reads as  
\begin{eqnarray}
 \!\!\! \bar{\sigma}_{xx} &=& \!\!\!
\frac{e^2}{h} \lim_{\eta \to 0^+} \eta^2 \!\! 
\int \! d^2 r \, r^2 \, 
\text{Tr}\left[ G(0,{\bf r},i\eta) G({\bf r},0,-i\eta)\right],
\label{cond_sigma_model:def}
\end{eqnarray}%
where the Green's function is given by Eq.~(\ref{Green:eq}). 
The conductivity $\bar{\sigma}_{xx} $ can be rewritten 
as\cite{Ludwig:cikk,McKane:cikk}
\begin{subequations}
\begin{eqnarray}
\bar{\sigma}_{xx} &=& 
-\frac{8 e^2}{h} \lim_{\eta \to 0^+} \eta^2 \, 
{\frac{\partial}{\partial q^2} 
	\,\rule[-1.6ex]{.2pt}{4ex}\;
	\raisebox{-1.5ex}{$\scriptstyle q=0$}}
K({\bf q},\eta), 
\label{sigma1:eq} \\[2ex]
\!\!\!\!\!\! K({\bf q},\eta) \!\!\! &=& \!\!\! 
\frac{1}{2}\!\! \int \! d^2 r \, e^{i{\bf q r}} \, 
\text{Tr}\left[ G(0,{\bf r},i\eta) G({\bf r},0,-i\eta)\right], 
\label{K:def}
\end{eqnarray}%
\end{subequations}%
and the trace is taken over the `pseudo spin space'. 

Using the Green's function given by Eq.~(\ref{Green:eq}) and 
performing the trace over the `pseudo spin space' in (\ref{K:def}) 
we find
\begin{subequations}
\begin{eqnarray}
K({\bf q},\eta) \!\!\! &=& \!\!\! 
\int \frac{d^2 k}{{\left(2\pi \right)}^2} \, 
\frac{\mbox{\boldmath $\Omega$}_+\mbox{\boldmath $\Omega$}_- +\eta^2}
{\left(\mbox{\boldmath $\Omega$}_+^2 +\eta^2\right)
\left(\mbox{\boldmath $\Omega$}_-^2 +\eta^2\right)}, 
\label{K1:eq} \\[2ex]
\mbox{\boldmath $\Omega$}_\pm &=& 
\mbox{\boldmath $\Omega$}({\bf k} \pm {\bf q}/2). 
\end{eqnarray}%
\label{K1-tot:eq}%
\end{subequations}%
After a simple algebra the integrand $I$ 
in Eq.~(\ref{K1:eq}) can be calculated explicitly:   
\begin{subequations}
\begin{eqnarray}
I \!\! &=& \!\!   
\frac{g^2 {\left[{\left( k^2 \! + \! q^2/4 \right)}^2 
\!\! - {\left({\bf k q}\right)}^2\right]}^{J/2}
\!\!\!\! \cos (J \Phi) 
\! + \! \eta^2}{\Sigma_+ \, \Sigma_-}, \\[2ex]
\Sigma_\pm &=& g^2 {\left(k^2 +\frac{q^2}{4} 
\pm {\bf kq} \right)}^J +\eta^2 ,\\[1ex]
\cos \Phi &=& \frac{k^2 - q^2/4}
{\sqrt{{\left( k^2 + q^2/4 \right)}^2 - {\left({\bf k q}\right)}^2}}.  
\end{eqnarray}%
\end{subequations}%
Here $\Phi$ is the angle between vector ${\bf k}+{\bf q}/2$ 
and ${\bf k}-{\bf q}/2$.
Expanding $I$ for small $q$ we obtain 
\begin{widetext}
\begin{eqnarray}
I &\approx& \frac{1}{g^2 k^{2J}+\eta^2} 
- q^2\, g^2 J k^{2J-2}\, \frac{J\left( g^2 k^{2J} +3 \eta^2 \right)
- \left[g^2 \left(J+1\right) k^{2J} - \left(J-1\right)\eta^2 \right]
\cos\left(2 \varphi -2 \gamma \right)}
{4{\left(g^2 k^{2J}+ \eta^2\right)}^3 },
\label{Integrand:eq}
\end{eqnarray}%
\end{widetext}%
where $\varphi$ and $\gamma$ are the polar angle of ${\bf k}$ and
${\bf q}$, i.e., ${\bf k}=k (\cos \varphi, \sin \varphi)$ and 
${\bf q}=q (\cos \gamma, \sin \gamma)$. As can be seen below $\gamma$
will be dropped out in the final result.  

Integrating the integrand $I$ in (\ref{Integrand:eq}) over ${\bf k}$ 
and taking the derivation with respect to $q^2$ 
we have the following simple result: 
\begin{equation}
{\frac{\partial}{\partial q^2}	
	\,\rule[-1.6ex]{.2pt}{3ex}\;
	\raisebox{-1.5ex}{$\scriptstyle q=0$}} K({\bf q},\eta) 
= -\frac{J}{8 \pi \eta^2},
\end{equation}
where $\gamma$ has been dropped out 
after the integration over $\varphi$. Similarly, the first term
disappears in $I$ since it is independent of $q$.
Now using Eq.~(\ref{sigma1:eq}) the conductivity (per valley per spin)
becomes
\begin{equation}
\bar{\sigma}_{xx} = \frac{J}{\pi}\, \frac{e^2}{h}.
\label{sigma_bar_per-valley_per-spin:eq}
\end{equation}

{\em Discussion:} 
We calculated the minimal conductivity for perfect single ($J=1$) and bilayer
($J=2$) graphene using the Kubo formula and the 
definition (\ref{cond_sigma_model:def}). 
Taking into account the two valleys ($n_v=2$) and the electron spin 
($n_s=2$) we find from Eqs.~(\ref{sigma__per-valley_per-spin:eq})
and (\ref{sigma_bar_per-valley_per-spin:eq}) that the minimal
conductivities take the universal values:  
\begin{subequations}
\begin{eqnarray}
\sigma^{\text min}_{xx} &=& \frac{J\pi}{2}\, \frac{e^2}{h}, 
\label{cond_min-1:eq}  \\
\bar{\sigma}^{\text min}_{xx} &=&  \frac{4 J}{\pi}\, \frac{e^2}{h}.
\label{cond_min-2:eq} 
\end{eqnarray}%
\end{subequations}%
Note that these results are independent of the constant $g$ appeared
in the Hamiltonian (\ref{HJ-1:eq}).
Just like in the case of single layer graphene we also find different
values for the conductivity when the two methods in
Ref.~\onlinecite{Ludwig:cikk} is extended to bilayer graphene. 
Moreover,  the minimal conductivity in perfect bilayer graphene 
calculated from Eqs.~(\ref{Kubo_cond:eq}) and (\ref{cond_sigma_model:def}) 
is doubled compared to that in single layer graphene. 
One can also see that for perfect single layer graphene our result
(\ref{cond_min-2:eq}) agrees with that obtained in
Refs.~\onlinecite{Gusynin-Hall_sxx:cikk,Peres-1:cikk,Katsnelson:cikk,Tworzydlo:cikk}. 

As pointed out by Katsnelson\cite{Katsnelson:cikk,Katsnelson-bilayer:cikk}, 
the reason for the different results (\ref{cond_min-1:eq}) and
(\ref{cond_min-2:eq}) is related to the `trembling' or
oscillatory motion of the center of a free wave
packet called {\em Zitterbewegung} 
which has recently been reconsidered in Ref.~\onlinecite{sajat_Zitter:cikk}. 
The Zitterbewegung induces an intrinsic disorder in  perfect graphenes
which has indirectly been confirmed by Tworzydlo et al.\
\cite{Tworzydlo:cikk} by calculating the Fano factor of the shot
noise in single layer graphene, and they found that 
it has the same value ($1/3$) as in disordered metals. 

In our calculation the energy dispersion of bilayer graphenes was
assumed to be purely parabolic in ${\bf k}$ (see Eq.~(\ref{HJ-1:eq})). 
Koshino and Ando studied the conductivity when the energy dispersion includes 
both k-linear and k-square terms\cite{Ando-bilayer:cikk} which results
in a trigonally warped Fermi surface\cite{Ed-Falko:cikk}.
To clarify the role of the k-linear term it would be interesting to  
extend our work along this line. 

Similarly, extending the theoretical efforts to study multi-layer graphenes 
would be desirable in order to have better understanding of the 
unusual transport in graphenes. 
Recently, the electronic states in stacks of
graphene layers have been studied\cite{Guinea_multi-layer:cikkek}. 
However, it is not clear whether in the case of trilayers
the Hamiltonian can be described simple by Eq.~(\ref{HJ-1:eq}) with $J=3$. 

The single and bilayer graphene are zero-gap semiconductors but this
gap can open up due to impurities or the  presence of interaction. 
Then, graphenes mapped to the massless Dirac fermion gas transform to
random mass Dirac fermion systems. 
In this case, to calculate the conductivity it is useful to start from the
definition (\ref{cond_sigma_model:def}) as it was demonstrated by
Ludwig et al.~\cite{Ludwig:cikk}. 
Recently, Gusynin et al.\ proposed theoretically that microwaves 
could be an experimental tool to understand the role of the impurities and 
the presence of interaction in graphene systems\cite{Gusynin-microwave:cikk}.

In summary, our work shows that the physical explanation of the existence of 
the minimal conductivity in graphene systems 
still remains a theoretical challenge in the future. 

We gratefully acknowledge discussions
with C. W. J. Beenakker, E. McCann, V. Fal'ko, B. Gy{\"o}rffy, B. Nikoli\'c, 
and A. Pir{\'o}th.
This work is supported by E.~C.\ Contract No.~MRTN-CT-2003-504574.


\end{document}